# Nonlinear transmission through a tapered fiber in rubidium vapor

S. M. Hendrickson,[1,2*] T. B. Pittman[1] and J. D. Franson[1]

[1]*Department of Physics, University of Maryland Baltimore County, Baltimore, MD 21250 USA*

[2]*Department of Electrical and Computer Engineering, John Hopkins University, Baltimore, MD 21218 USA*

[*]*Corresponding author: hendrix@jhu.edu*

**Abstract:** Sub-wavelength diameter tapered optical fibers surrounded by rubidium vapor can undergo a substantial decrease in transmission at high atomic densities due to the accumulation of rubidium atoms on the surface of the fiber. Here we demonstrate the ability to control these changes in transmission using light guided within the taper. We observe transmission through a tapered fiber that is a nonlinear function of the incident power. This effect can also allow a strong control beam to change the transmission of a weak probe beam.

## 1. Introduction

Tapered optical fibers (TOFs) with diameters less than the wavelength of the guided light have been well characterized [1] and shown to have a wide range of applications including optical sensing [2], microtoroidal coupling [3] and the observation of various linear and nonlinear optical effects [4-11]. Because a portion of the power is guided in the evanescent field [12], TOFs experience significant changes in transmission when the silica-air boundary is contaminated. For this reason care is taken to isolate TOFs from dust and other contaminants.

In experiments that involve atomic media, the degradation of the surface of the fiber can be more difficult to prevent because atoms tend to accumulate. This is especially troublesome with rubidium and special techniques are often needed to prevent a loss in transmission. It has been suggested that coating the surface of the TOF with a substance such as organosilane [13] may prevent the degradation but has not yet been demonstrated on fiber devices of this type. Under appropriate conditions, sufficient heating of the taper may reduce the accumulation of rubidium [11]. In addition, there is evidence to suggest that the atoms could be desorbed using light-induced atomic desorption (LIAD), in which photons provide the energy required to remove the atoms from the surface. LIAD has been observed on various substrates coated with organic materials [14] and paraffin [15] as well as on uncoated surfaces [16].

In this work we demonstrate a non-resonant effect in which atomic rubidium is removed from the surface of a TOF using low-power light propagating through the fiber. This results in transmission through the TOF that is a nonlinear function of the incident power. These results are believed to be due to a combination of LIAD [17, 18, 19] and heating of the tapered region. In any event, these effects represent a nonlinear relationship between the input and output powers of this system rather than a true nonlinear optical effect.

## 2. Experiment

The setups shown in Fig. 1 were used to investigate the transmission through the TOF as a function of input power.  The main component of both experiments was a TOF placed in a vacuum chamber.  The TOF was fabricated on a stainless steel mount using the flame-brush technique [1] with an oxygen-butane flame.  The TOF was secured to the mount with Viton® washers to allow the entire setup to be transferred into the vacuum chamber.  The fiber pulling was performed with computer controlled micro-positioners while the transmission through the fiber was monitored to determine when the taper diameter was in the single-mode regime.  The taper waist diameter ranged from 450 nm to 650 nm over a length of approximately 2 mm.  The TOF was connected to optical fibers outside the vacuum chamber through custom-machined Teflon® feedthroughs [20] and connected to standard fiber using gel-splices.  The vacuum system was first baked for 24 hours at 160°C to reduce subsequent outgassing at lower temperatures.  The background pressure after bake-out was typically $10^{-8}$ torr.  A valve could then be opened to release rubidium into the chamber.  During experiments the primary chamber was kept at 125°C and the rubidium source at 145°C.

A small radiative heating element consisting of a copper rod in contact with a black anodized aluminum radiating element was placed a few millimeters above the TOF in the vacuum chamber.  This rod was heated to 155°C throughout the baking process and left at that temperature when the remainder of the system was cooled.  This prevented the long-term accumulation of large amounts of rubidium on the taper, which would reduce the transmission to the point where it could not be recovered using our optical technique.

*A. Self-modulation of transmission*

A plot of the typical relation between input power and percent transmission is shown in Fig. 2.  This data was taken using the setup shown in Fig. 1(a) with a probe frequency 100 GHz detuned from the D$_2$ line of rubidium using a frequency-stabilized diode laser. The plot shows that at low input power the transmission through the TOF is negligible. As the power is increased there is a rapid increase in the percentage of transmitted light. The transmission then levels off and slowly approaches the transmission properties exhibited by the taper in the absence of rubidium vapor. Similar dependence on input power was observed at wavelengths throughout the range of our tunable laser (about 760-780 nm) as well as at lower wavelengths such as 650 nm. The data shown in Fig. 2 were taken with 15 seconds of settling time for each point. We observed a slow increase with the value of each point if settling times were longer, particularly at the higher range of the figure. As a result of this settling time the transmission exhibited hysteresis upon reduction of the input power [21].

*B. Control of a weak probe beam*

The ability of a guided beam to increase the transmission of the TOF allows the use of a strong control beam to modulate the transmission of a weak probe beam. To demonstrate this, we used the setup shown in Fig. 1(b) with a counter-propagating control beam at 770 nm and a weak (~ 1 µW) probe beam at 780 nm. The back-reflections from the control beam were then filtered out using a spectral filter and polarization analyzers as needed.

Figure 3 shows that a control beam with power on the order of microwatts can be used to control the transmission of a counter-propagating probe beam of a different wavelength and much lower power. The control beam was able to modulate the probe beam by 16 dB. In principle, this result suggests that this effect could be used to construct an all-optical switch or modulator.

It should be noted that the curves in Figs. 2 and 3 are not expected to have the same form, since the power lost by the probe beam in desorbing the atoms in Fig. 2 reduces the transmission of the probe, which is not the case in Fig. 3.

*C. Atomic spectroscopy of the Rb $D_2$ line*

The effects described above may be of practical use in spectroscopy using tapered optical fibers in atomic vapors when the density is sufficiently high that there is a significant accumulation of atoms on the fiber. In order to illustrate this, we have measured the spectrum of the well-known $D_2$ line in rubidium [11, 22], as shown in Fig. 4. The laser frequency was scanned over the region corresponding to the absorption lines of the two hyperfine ground-state levels of naturally-occurring $^{85}$Rb (F = 2 and 3) and $^{87}$Rb (F = 1 and 2), which results in four dips in the transmission curve at wavelengths near 780 nm. (Two of the lines are not well resolved here due to Doppler, collisional and time-of-flight broadening.)

Figures 4(a) and 4(b) were obtained using only a single probe beam with no control beam present. Figure 4(a) was measured with a probe intensity of 0.95 µW and shows the four absorption lines mentioned above, while the absorption lines have been almost completely eliminated in the data of Fig. 4(b), which were obtained with a probe intensity that was approximately a factor of five larger than in Fig. 4(a). This reduction in the absorption at higher powers is a saturation effect [11, 23, 24] that is caused primarily by a depopulation of the hyperfine ground state that is currently on resonance with the laser due to optical pumping into the other hyperfine ground state that is not currently on resonance. At these intensities, this effect is much larger than the effects of power broadening [25], and the peaks in the data of Fig. 4 are not significantly broadened as a result.

The data of Fig. 4 show several interesting features. Ordinarily, a factor of five increase in the probe intensity would reduce the saturated absorption signal by at most a factor of five, while a comparison of Figs. 4(a) and 4(b) shows substantially more saturation than that. Increasing the intensity by a factor of five desorbs atoms from the surface of the fiber and increases the transmission, so the average intensity in the tapered region increases by much more than a factor of five, which is why the saturation effects are more pronounced here than would ordinarily be the case.

Figures 4(c) and 4(d) were obtained using a control beam (770 nm) with an intensity of 49 µW but with two different probe intensities. Based on the results of Figs. 2 and 3, the control beam would be expected to desorb most of the rubidium atoms from the surface of the tapered fiber. A comparison of Figs. 4(a) and 4(c) shows that spectroscopic measurements of this kind can be made at much lower intensities in the presence of a control beam, since there would have been almost no transmission at the probe intensity (8 nW) used in Fig. 4(c) without the control beam.

A comparison of Figs. 4(a) and 4(d), which correspond to comparable probe intensities, show that the presence of the control beam allows saturated absorption to occur at lower probe intensities than would be the case without the control beam. Although a probe intensity of 1.1 µW is not sufficient to produce total saturation even in the presence of a strong control beam, the saturation effects are clearly much larger in Fig. 4(d) than in Fig. 4(a). Once again, the control beam desorbs atoms from the surface of the fiber, which reduces the amount of loss and increases the average intensity for a given input power. The increased intensity is then responsible for the increased saturation effects [26].

## 3. Conclusions

All of the effects described above were greatly reduced when the rubidium oven was cooled back down to room temperature to reduce the rubidium density. This suggests that these effects are indeed due to rubidium on the surface of the tapered region.

We have shown that the transmission through a tapered optical fiber in rubidium vapor is a nonlinear function of input power. These effects can be used as an optical method to prevent the build-up of rubidium on TOFs. Control of a weak probe beam by a non-resonant control beam has also been demonstrated and may be of practical use as a switch or modulator. Preliminary tests of atomic spectroscopy in a TOF using this technique have been carried out.

**Acknowledgements**

This work was supported in part by the Intelligence Advanced Research Projects Activity (IARPA) under Army Research Office (ARO) contract W911NF-05-1-0397, and the National Science Foundation under grant No. 0652560.

# LIST OF FIGURES

## A. Self-Modulation

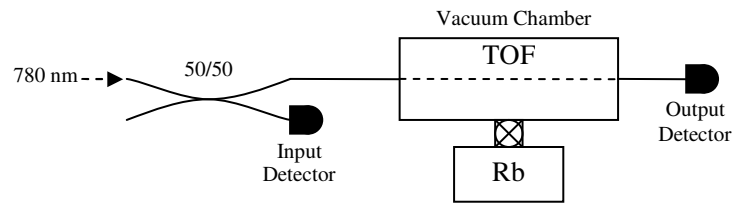

## B. Control-Probe Modulation

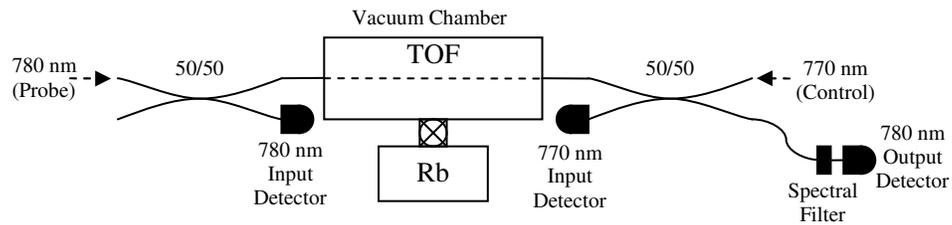

Fig. 1. The experimental setups used to investigate the nonlinear transmission through the TOF. (a) Self-modulation of transmission (b) Control-probe modulation of transmission

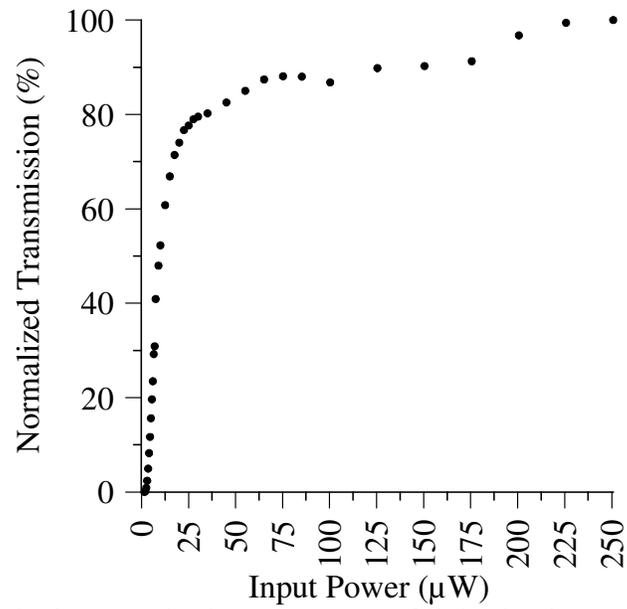

Fig. 2. A plot showing the increase in the percentage of light that is transmitted through a TOF as a function of the input power. Transmission is normalized to the maximum transmission at high power levels.

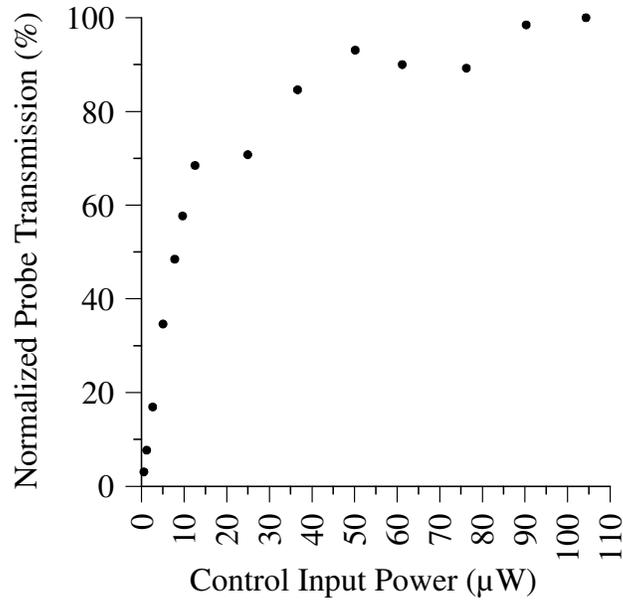

Fig. 3. A plot showing the increase in the transmission of a weak probe beam (~1μW; 780 nm) as a function of the input power of a stronger counter-propagating control beam (770 nm). Transmission is normalized to the maximum transmission at high control beam power levels. The stronger beam has been removed using a tunable wavelength filter as shown in Fig. 1(b) and the input power of the probe beam remained constant.

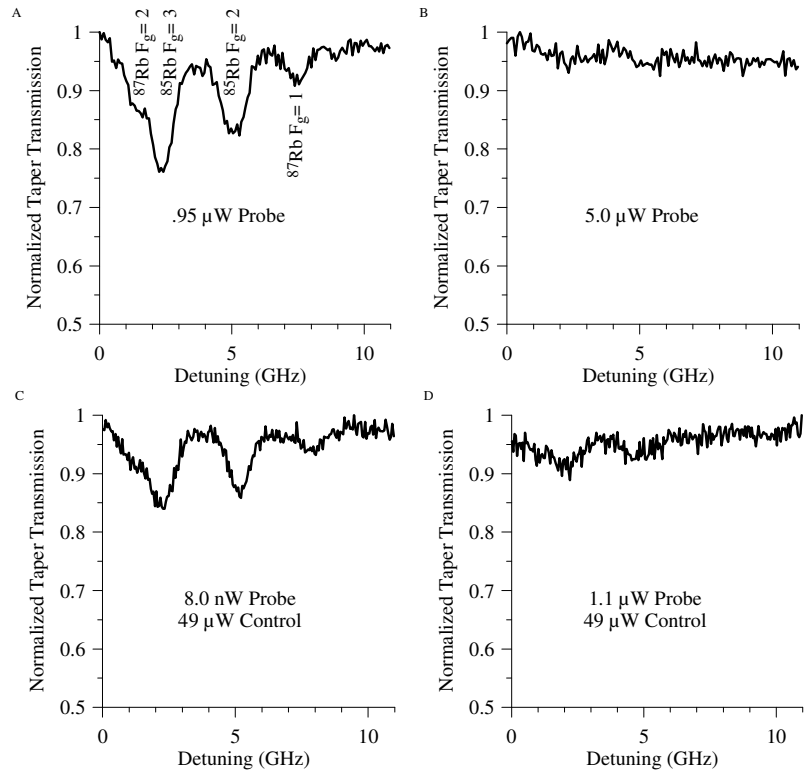

Fig. 4. Plots showing the transmission through the tapered fiber as a function of frequency detuning near the $D_2$ line of natural rubidium. These plots are normalized to the maximum value measured in each laser scan for comparison purposes. (a) No control beam, .95 μW of probe beam at 780 nm (b) No control beam, 5.0 μW of probe beam at 780 nm (c) With 49μW of control beam power applied at a wavelength of 770 nm, and a probe beam with 8.0 nW of power at 780 nm (d) A control beam with 49 μW of power at 770 nm, and a probe beam with 1.1 μW of power at 780 nm. The reduction in the depths of the resonant features in (b) and (d) are due to the saturation of the atomic transitions as described in the text.